\documentclass[nofootinbib,floatfix,preprint,showpacs,showkeys,amsmath,amssymb]{revtex4}

\usepackage{graphicx}% Include figure files
\usepackage{dcolumn}% Align table columns on decimal point
\usepackage{bm}% bold math
\def\pslash{p\!\!\!\slash }
\begin{document}

%\preprint{}

\title{Finite-width effects on Delta baryons in QCD Sum Rules}% Force line breaks with \\

 \author{G. Erkol}
 \email{erkol@th.phys.titech.ac.jp}
 \author{M. Oka}%
  \email{oka@th.phys.titech.ac.jp}
 \affiliation{%
 Department of Physics\\ Tokyo Institute of Technology\\ Meguro, Tokyo 152-8551 Japan
 }%

\date{\today}% It is always \today, today,

\begin{abstract}
The quantum chromodynamics (QCD) sum rules for the Delta baryons are analyzed by taking into account the finite-width effects, through explicit utilization of the Breit-Wigner shape. We apply a Monte-Carlo based analysis to the `traditional' and the parity-projected sum rules. The first Delta excitation state is also considered as a sub-continuum resonance and the widths are calculated using the mass values as input. 
\end{abstract}

%Use showkeys class option if keyword
\maketitle

%%%%%%%%%%%%%%%%%%% Section 1 %%%%%%%%%%%%%%%%%%%%%%%%%%%%
\section{Introduction}
The Delta baryons, with spin and isospin $3/2$, have long been a subject of study~\cite{Ioffe:1981kw,Belyaev:1982sa,Reinders:1984sr,Kodama:1994pn,Hwang:1994vp,Lee:1997ix,Lee:2006bu,Erkol:2007sj} in QCD Sum Rules (QCDSR), which are a well-established method that links the hadron degrees of freedom with the underlying QCD parameters~\cite{Shifman:1978bx,Shifman:1978by}. An important issue in the analysis of the decuplet baryons as well as of other excited hadrons is their finite widths. Conventionally, the method is to apply the zero-width approximation, where the Breit-Wigner contributions to the imaginary parts of the correlation functions are represented by $\delta$-functions~\cite{Shifman:1978bx,Shifman:1978by}. However, one cannot automatically assume that the contributions of the finite-width effects are negligible. Therefore, it is necessary to go beyond analytical arguments and investigate numerically the finite-width effects on the hadron spectrum in QCDSR.

The finite-width contributions within QCDSR have been considered before in the context of the $\rho$-meson~\cite{Leupold:1997dg} and the $\rho$--$\omega$ mixing~\cite{Iqbal:1995gp,Iqbal:1995qc}. Our aim in this work is to make a reanalysis of the Delta sum rules by taking into the account the finite-width effects. For this purpose, we first consider the `traditional' sum rules and in order to implement the finite-width effects on the $\Delta(1232)$ state, we use the method developed in Ref.~\cite{Elias:1997ya}, which was then applied to the first pion excitation state, $\pi(1300)$. Our second aim in this paper is to include the first positive- and negative-parity Delta excitation states, $\Delta(1600)$ and $\Delta(1700)$ respectively, within QCDSR analysis as sub-continuum resonances and to extract information about the parameters that characterize these resonances. For this purpose we resort to the parity-projected sum rules in order to separate the positive- and the negative-parity contributions~\cite{Jido:1996zw,Jido:1996ia,Lee:2006bu}.  

For our numerical procedure, we use the Monte-Carlo based analysis introduced in Ref.~\cite{Leinweber:1995fn}. This method provides a more systematic treatment of uncertainties in QCDSR. On the other hand, we expect that a Monte Carlo based analysis can better resolve the discrepancies in the hadron spectrum due to finite-width contributions. We have organized our paper as follows: in Sec.~\ref{sect2}, we present the formulation of the traditional Delta sum rules including the finite-width contributions. In Sec.~\ref{sect3}, we consider the parity-projected sum rules and compare the results with those from the traditional ones. Finally, we arrive at our conclusions in Sec.~\ref{sect4}.
 
%%%%%%%%%%%%%%%%%%% Section 2 %%%%%%%%%%%%%%%%%%%%%%%%%%%%
\section{Traditional Sum Rules} \label{sect2}
%%%%%%%%%%%%%%%%%%% Subsection 2-a %%%%%%%%%%%%%%%%%%%%%%%%%%%%
\subsection{Construction of the Traditional Sum Rules}
We begin our analysis by considering the correlation function
	\begin{equation}\label{cor1}
	   \Pi_{\mu\nu} = i\int d^4 x~ e^{i p\cdot x}\, \Big \langle 0\Big|{\cal T}[\eta_\mu(x)\bar{\eta}_\nu(0)]\Big |0\Big\rangle,	
	\end{equation}
where $\eta$ is the Delta interpolating field:
	\begin{equation}\label{intf}
		\eta_\mu(x) = \epsilon_{abc}[u_a^T(x) C\gamma_\mu u_b(x)] u_c(x) .
	\end{equation}
Here $a,b,c$ are the color indices, $T$ denotes transposition and $C=i\gamma^2\gamma^0$. The QCDSR are derived by calculating the correlation function in Eq.~(\ref{cor1}) in the deep Euclidian region using Operator Product Expansion (OPE) and matching this to a phenomenological ansatz. The correlation function in Eq.~(\ref{cor1}) has the following form:
	\begin{equation}\label{ldst}
	\Pi_{\mu\nu}(p)=\Pi_1(p^2)g_{\mu\nu}+\Pi_2(p^2)g_{\mu\nu}\pslash+\ldots,
	\end{equation}
where the ellipsis represents the Lorentz-Dirac structures other than $g_{\mu\nu}$ and $g_{\mu\nu}\pslash$. In principle, one can obtain the sum rules at different Lorentz-Dirac structures. Here, we are interested in the $g_{\mu\nu}$ and the $g_{\mu\nu}\pslash$ structures, which are contributed by only the $J=3/2$ particles (see e.g. Refs.~\cite{Ioffe:1981kw,Hwang:1994vp} for a more explicit separation of the spin components).

The analyticity of the correlation function allows us to write the phenomenological side of the sum rules in terms of a dispersion relation of the form
	\begin{equation}\label{phenside}
		\text{Re}\,\Pi_{\mu\nu}(q^2)=\frac{1}{\pi}\int^\infty_0 \frac{\text{Im}\,\Pi_{\mu\nu}(s)}{(s-p^2)}\,ds.
	\end{equation}
The ground-state hadron contribution is singled out by utilizing the zero-width approximation, where the hadronic contributions from the Breit-Wigner form to the imaginary part of the correlation function is proportional to the $\delta$-function:
	\begin{align}\label{impart}
	\begin{split}	
		\lim_{\Gamma\rightarrow0}\text{Im}\Big[\frac{-1}{s-m^2+i\, m\,\Gamma}\Big] =\pi\delta(s-m^2),\\
		\text{Im}\,\Pi_{\mu\nu}(s)=\pi\sum_X\, \delta(s-m_X^2)\langle 0 |\eta_\mu|X\rangle \langle X |\eta_\nu|0\rangle.
	\end{split} 
	\end{align}
One then expresses the correlation function in Eq.~(\ref{cor1}) as a sharp resonance plus a continuum after Borel transformation:
	\begin{align}\label{phenside2}
	\begin{split}
		&\Pi_1(M^2)=\tilde{\lambda}_\Delta^2\, m\,e^{-m^2/M^2}+\frac{1}{\pi}\int_{w^2}^\infty\,\text{Im} \,\Pi_1\, e^{-s/M^2}\,ds,\\
		&\Pi_2(M^2)=\tilde{\lambda}_\Delta^2 \,e^{-m^2/M^2}+\frac{1}{\pi}\int_{w^2}^\infty\,\text{Im}\,\Pi_2\, e^{-s/M^2}\,ds,
	\end{split}
	\end{align}
at the structures $g_{\mu\nu}$ and $g_{\mu\nu}\pslash$, respectively. Here, $M$ is the Borel mass, $w$ is the continuum threshold and the matrix element of the current $\eta^\mu$ between the vacuum and the Delta state is defined as 
	\begin{equation}~
		\label{overlapD} \langle 0 | \eta^\mu | \Delta(p,s) \rangle= \tilde{\lambda}_\Delta \upsilon^\mu(p,s), 
	\end{equation}
where $\tilde{\lambda}_\Delta$ is the overlap amplitude and $\upsilon^\mu(p,s)$ is the Rarita-Schwinger spin vector of the Delta. We also make use of the Rarita-Schwinger spin sum, which is
	\begin{equation}~\label{RSss}
		\sum_s \upsilon^\mu(p,s) \bar{\upsilon}^\nu(p,s)=-\Big(g^{\mu\nu}-\frac{1}{3}\gamma^\mu \gamma^\nu-\frac{p^\mu \gamma^\nu-p^\nu\gamma^\mu}{3\,m}-\frac{2\,p^\mu p^\nu}{3\,m^2}\Big)(\pslash+m). 
	\end{equation}

For the sake of completeness, we first present the sum rules in momentum space:
	\begin{align}\label{momsp}
		\begin{split}
			&(g_{\mu\nu}):\quad \frac{4}{3}\,a_q\,p^2\,\ln(-p^2)-\frac{2}{3}\, m_0^2\, a_q\, \ln(-p^2)-\frac{1}{18p^2}\,a_q\,b = \lambda^2\,m\, \int^{\infty}_0 \frac{I(m,\Gamma)}{(p^2-s)} \,ds,\\
			&(g_{\mu\nu}\pslash):\quad \frac{p^4}{10}\,\ln(-p^2)-\frac{5}{72}\, b\,\ln(-p^2)+\frac{4}{3p^2}\,\kappa\,a_q^2+\frac{7}{9p^4} \,m_0^2\,a_q^2= \lambda^2 \, \int^{\infty}_0 \frac{I(m,\Gamma)}{(p^2-s)} \,ds.
		\end{split}
	\end{align}
Upon Borel transformation, the chiral-odd sum rule (OSR) at the structure $g_{\mu\nu}$ is given as
	\begin{equation}\label{cosumrule}
		\frac{4}{3}a\,E_1\,L^{16/27}M^4-\frac{2}{3}E_0\,m_0^2\, a\, L^{2/27}M^2 -\frac{1}{18} a\, b\, L^{16/27}=\lambda^{2}\,m \int^{w^2}_0 I(m,\Gamma)\,e^{-s/M^2}\,ds ,
	\end{equation}
and the chiral-even sum rule (ESR) at the structure $g_{\mu\nu}\pslash$ for the Delta is given as
	\begin{align}\label{cesumrule}
		\begin{split}	
		&\frac{1}{5}E_2\,L^{4/27}M^6-\frac{5}{72}b\,E_0\,L^{4/27}M^2 +\frac{4}{3}\kappa\, a^2\,L^{28/27}-\frac{7}{9\,M^2} m_0^2\,a^2\,L^{17/27}\\
		&\quad=\lambda^{2}\,\int^{w^2}_0 I(m,\Gamma)\,e^{-s/M^2}\,ds,
		\end{split}
	\end{align}
where $\lambda =(2\pi)^2\tilde{\lambda}_\Delta$~\cite{Ioffe:1981kw,Hwang:1994vp,Lee:1997ix}. To include the finite-width effects, we replace the $\delta$-functions in Eq.~(\ref{impart}) by a Breit-Wigner form~\cite{Dominguez:1984eh}
	\begin{equation}\label{BWf}
		I(m,\Gamma)=\frac{1}{\pi}\frac{m\Gamma}{(s-m^2)^2+m^2\Gamma^2},
	\end{equation} 
where $\Gamma$ is the width of the resonance state. In the sum rules above we have defined the quark condensate $a=-(2\pi)^2\langle\bar{q}q\rangle$, the gluon condensate $b=\left\langle g_c^2 G^2\right\rangle$, and the quark-gluon mixed condensate $\left\langle\bar{q}g_c\bm{\sigma}\cdot\bm{G} q\right\rangle=m_0^2 \langle\bar{q}q\rangle$ with the QCD coupling-constant squared $g_c^2=4\pi\alpha_s$. The four-quark condensate is parameterized as $\langle(\bar{q}q)^2 \rangle\equiv\kappa \langle\bar{q}q\rangle^2$. The corrections that come from the anomalous dimensions of various operators are included with the factors $L=\log(M^2/\Lambda_{QCD}^2)/\log(\mu^2/\Lambda_{QCD}^2)$, where $\mu=500$~MeV is the renormalization scale and $\Lambda_{QCD}$ is the QCD scale parameter. The perturbative corrections are taken into account with the factors~\cite{Chung:1984gr}
	\begin{align}\label{betas}
		\begin{split}
		&\text{dimension-three  ($a$):}\quad 1+\Big(\frac{11}{9}-\frac{4}{3}\gamma_E\Big) \frac{\alpha_s}{\pi},\\ 
		&\text{dimension-five ($m_0^2\,a$):}\quad 1+\Big(\frac{17}{27}-\frac{5}{6}\gamma_E\Big) \frac{\alpha_s}{\pi},
		\end{split}
	\end{align}
in OSR, and with
	\begin{align}\label{betas2}
		\begin{split}
		&\text{dimension-zero ($I$):}\quad 1+\Big(\frac{539}{90}-\frac{1}{3}\gamma_E\Big) \frac{\alpha_s}{\pi},\\
		&\text{dimension-six ($a^2$):}\quad 1-\Big(\frac{113}{108}+\frac{22}{3}\gamma_E\Big) \frac{\alpha_s}{\pi},
		\end{split}
	\end{align}
in ESR, where $\gamma_E\simeq 0.58$ is the Euler constant. These corrections give large contributions to ESR (leading-order correction is $\sim 70\%$ for $\alpha_s/\pi\simeq 0.12$ at the scale of 1~GeV$^2$) whereas their contributions to OSR is small and therefore can be safely neglected. The continuum contributions are represented by the factors $E_n(x)=1-e^{-x}\sum_n x^n/n!$, with $x=w^2/M^2$.

%%%%%%%%%%%%%%%%%%% Subsection 2-b %%%%%%%%%%%%%%%%%%%%%%%%%%%%
\subsection{Numerical Analysis of the Traditional Sum Rules}
In order to obtain an analytic approximation for the integral over the Breit-Wigner shape, we first express the imaginary part of the Breit-Wigner shape as a Riemann sum of unit-area rectangular pulses $P_m(s,\Gamma)$~\cite{Elias:1997ya}:
	\begin{align}\label{unitpulse}
		\begin{split}
		 P_m(s,\Gamma)&\equiv[\Theta(s-m^2+m\Gamma) -\Theta(s-m^2-m\Gamma)]/2m\Gamma,\\
		 \frac{m\Gamma}{(s-m^2)^2+m^2\Gamma^2}&= \lim_{n\rightarrow \infty}\frac{2}{n}\sum^n_{j=1}\sqrt{\frac{n}{j-f}-1} \,\,P_m\Big(s,\sqrt{\frac{n}{j-f}-1}\,\,\Gamma\Big),\quad 0\leq f \leq 1,
		\end{split}
	\end{align}
where an approximation for the integral over Breit-Wigner shape can be obtained by using (for $w^2>m^2+m\,\Gamma$)
	\begin{align}\label{ibw1}
		\begin{split}
		\int^{w^2}_0 P_m(s,\Gamma)\,e^{-s/M^2}\, ds&=e^{-m^2/M^2}G(m,\Gamma,M^2),\\
		G(m,\Gamma,M^2)&\equiv\frac{M^2}{m\Gamma}\sinh \Big(\frac{m\Gamma}{M^2}\Big).
		\end{split}
	\end{align}
In order to proceed, we have made a 4-pulse approximation by taking $n=4$ in Eq.~(\ref{unitpulse}) and choosing $f=0.7$ in order to make the area of the four pulses equal to the total area under the Breit-Wigner curve ($=\pi$)\footnote{In the $n\rightarrow\infty$ limit, Eq.~(\ref{unitpulse}) would be true for any choice of $f$ between 0 and 1}. One can then write the integral over the Breit-Wigner shape as
	\begin{equation}\label{fpa}
		\frac{1}{\pi}\int^{w^2}_0 \frac{m\Gamma}{(s-m^2)^2+m^2\Gamma^2}e^{-s/M^2} \,ds=e^{-m^2/M^2}\,W[m,\Gamma,M^2],
	\end{equation}
where
	\begin{align}
	\begin{split}	
		&W[m,\Gamma,M^2]=0.1592\,\Big[G(m,3.5119\,\Gamma,M^2) +G(m,1.4412\,\Gamma,M^2)\\ 
		&\quad+G(m,0.8597\,\Gamma,M^2)+G(m,0.4606\,\Gamma,M^2)\Big].
	\end{split}
	\end{align}
Note that $W\rightarrow 1$ in the narrow-resonance limit ($\Gamma\rightarrow 0$). This four-pulse approximation serves to reduce the numerical difficulties in evaluating the integrals over the Breit-Wigner shape, which extends to low-s region. We will further discuss this issue in Section~\ref{sect3}.

\begin{table}
	\addtolength{\tabcolsep}{4pt}
	\caption{The obtained values of the parameters representing $\Delta(1232)$ from a consideration of 1000 parameter sets and using OSR in Eq.~(\ref{cosumrule}) and ESR in Eq.~(\ref{cesumrule}). The values of the search parameters are given both in the zero-width limit ($\Gamma\rightarrow 0$) and with finite width ($\Gamma=0.12$~GeV). The fixed parameters are denoted by an asterisk in each case and the numbers inside brackets are experimental values as given by PDG~\cite{Yao:2006px}.} 
	\begin{tabular}{ccccccc}
		\hline\hline  Region (GeV) & $\Gamma$ (GeV) & $m$ (GeV) & $w$ (GeV) & $\lambda^2$ (GeV$^6$) & \\
		\hline   & $[0.118 \pm 0.002]$ & $[1.232]$ & & \\
		 ESR & $0~^\ast $ & $1.21 \pm 0.26$ & $1.65 \pm 0.25$ & $1.69 \pm 1.04$ \\[-1ex]
		 $1.15 \le M \le 1.35$& $0.12^\ast$ & $1.24 \pm 0.21$ & $1.60 \pm 0.23$ & $1.61 \pm 0.83$ \\[1ex]
		  & $0.28 \pm 0.09$ & $1.232~^\ast$ & $1.57 \pm 0.08$ & $1.34 \pm 0.29$ \\[-1ex]
		  & $0.27 \pm 0.09$ & $1.23 \pm 0.12~^\ast$ & $1.55 \pm 0.13$ & $1.36 \pm 0.37$ \\		
		\hline OSR & 0 & $1.43 \pm 0.12$& $1.65 \pm 0.22$& $2.39 \pm 0.94$  \\[-1ex]
		  $0.95 \le M \le 1.10$& $0.12~^\ast$ & $1.48 \pm 0.11$ & $1.72 \pm 0.21$ & $2.80 \pm 0.99$ \\[1ex]
		  & $0.16 \pm 0.06$ & $1.48~^\ast$ & $1.70 \pm 0.04$ & $2.69 \pm 0.39 $  \\[-1ex]
		  & $0.20 \pm 0.08$ & $1.48 \pm 0.15~^\ast$ & $1.76 \pm 0.20$ & $3.12 \pm 1.18$ \\[1ex]
		\hline\hline
	\end{tabular}
	\label{hadpar}
\end{table}
	
We determine the uncertainties in the extracted parameters via the Monte Carlo based analysis introduced in Ref.~\cite{Leinweber:1995fn}. In this analysis, randomly selected, Gaussianly distributed sets are generated from the uncertainties in the QCD input parameters. Here we use $a=0.52 \pm 0.05$~GeV$^3$, $b=1.2 \pm 0.6$~GeV$^4$, $m_0^2=0.72 \pm 0.08$~GeV$^2$, and $\Lambda_{QCD}=0.15 \pm 0.04$~GeV. The factorization violation in the four-quark operator is searched via the parameter $\kappa$, where we take $\kappa=2 \pm 1$ and $1 \leq \kappa \leq 4$; here $\langle \bar{q}q^2\rangle \geq \langle \bar{q}q\rangle^2$ is assumed via the cut-off at 1 (for a discussion on QCD parameters see {\it e.g.} Ref.~\cite{Leinweber:1995fn}). We use 1000 such configurations from which the uncertainty estimates in the extracted parameters are obtained. 

We begin our analysis by making a three-parameter fit to the left-hand sides (LHS) of the sum rules in Eq.~(\ref{cosumrule}) and Eq.~(\ref{cesumrule}), including $\lambda$, $m$ and $w$. The valid Borel regions are determined so that the highest-dimensional operator contributes no more than $10\%$ to the OPE side in order to warrant the OPE convergence, while the continuum contribution is less than $50\%$ of the phenomenological side, which provides pole domination. Note that, while the first criterion is rather straightforward, one does not initially have a complete control on the second, since the phenomenological parameters are determined from the fit and they are correlated. We use the following strategy: we first make the fits in a reasonably selected Borel region, which is then adjusted by trial and error according to the fit results, until the above criteria are satisfied. We also seek a region which best resolves the resonances and finally obtain an optimized Borel region. 

In Table~\ref{hadpar}, we present the obtained values of these parameters from a consideration of 1000 parameter sets. The values of the search parameters are given both in the zero-width limit ($\Gamma\rightarrow 0$) and with finite width ($\Gamma\rightarrow 0.12$~GeV)~\cite{Yao:2006px}. In the zero-width approximation, the Delta mass values we obtain are in agreement with those in Ref.~\cite{Lee:1997ix}. It is well-known that OSR performs better as compared to ESR even though OSR somewhat overestimates the Delta mass. We observe from the fitted parameter values in Table~\ref{hadpar} that the finite-width effects lead to a change of $\sim 3\%$ in the Delta mass whereas these effects change the continuum thresholds by 3--5\%. Moreover, the uncertainties in the values of the fit parameters from ESR are reduced by $\sim 20\%$ with the finite-width effects. We have also made a three-parameter fit including $\Gamma$, $\lambda$ and $w$ with the input $m=1.232$~GeV for ESR, and $m=1.48$~GeV for OSR. From ESR, we obtain an overestimated value for the width as $\Gamma=0.28 \pm 0.09$~GeV, which should be compared with the experimental result $\Gamma=0.118 \pm 0.002$~GeV. On the other hand, OSR successfully produces the experimental width value. Note that the overlap amplitude values we obtain from the Monte-Carlo fits are in agreement with the ones from earlier determinations: $\lambda^2=2.3 \pm 0.6$~GeV$^6$ from the traditional analysis of the sum rules~\cite{Belyaev:1982sa}, $\lambda^2=2.0 \pm 0.7$~GeV$^6$ from lattice QCD~\cite{Chu:1993cn} and $\lambda^2=1.8 \pm 0.5$~GeV$^6$ from instanton liquid model~\cite{Schafer:1993ra}.

The main conclusion of Table~\ref{hadpar} is that the mass determination is rather insensitive on the width variation. The masses in the zero-width approximation and with the finite-width effects agree within their uncertainties. At this point, it is also important to analyze the correlations between the phenomenological parameters, in particular between the mass and the width. The truncated OPE does not have enough information to allow a simultaneous determination of these two parameters. In order to extract one of them, we require the other as an input from experiment. However, given the approximate nature and limited accuracy of QCDSR (considering the truncated OPE, uncertainties in the condensate values and the continuum model), one needs to consider variations in all input parameters, including the mass and the width. For this purpose, we have allowed a variation of 10\% in the mass values by generating Gaussianly distributed sets from Monte Carlo, and studied the correlations by taking these sets as input. Our results are given in Table~\ref{hadpar}: we observe that a variation in the mass leads to larger uncertainties and deviations for the width in OSR, while this behavior is not evident for ESR. 

%%%%%%%%%%%%%%%%%%% Section 3 %%%%%%%%%%%%%%%%%%%%%%%%%%%%
\section{Parity-projected Sum Rules} \label{sect3}
%%%%%%%%%%%%%%%%%%% Subsection 3-a %%%%%%%%%%%%%%%%%%%%%%%%%%%%
\subsection{Construction of the Sum Rules}
Our next task in this analysis is to include the first Delta excitation on the phenomenological side by assuming that both the lowest-lying and the first-excitation Delta states reside below the continuum threshold and all the other subsequent excitations can be embedded in the continuum. One of the obstacles with this consideration is the identification of the first excitation, since the interpolating field in Eq.~(\ref{intf}) couples not only to the positive-parity but also to the negative-parity state~\cite{Jido:1996zw,Jido:1996ia}. If one is interested in the lowest-lying positive-parity resonance, the excited states can be regarded as parts of the continuum. However, when the first excitation is taken as a sub-continuum resonance and included as a pole on the phenomenological side, the question arises whether this pole actually represents the positive-parity $\Delta(1600)$ resonance or the negative-parity $\Delta(1700)$ resonance. The complexity of the excited-state spectrum may not allow us to separate the resonances of interest from the continuum by a proper choice of the threshold. Therefore, in the following, we apply parity projection to the correlation function in Eq.~(\ref{cor1}) in order to separate the positive- and the negative-parity contributions~\cite{Jido:1996zw,Jido:1996ia,Lee:2006bu} so that we can include the first excitation in our QCDSR analysis. For this purpose, we use the old-fashioned correlation function, which is
	\begin{equation}\label{cfof}
		\Pi^p_{\mu\nu} = i\int d^4 x~ e^{i p\cdot x}\, \theta(x_0) \Big \langle 0\Big|{\cal T}[\eta_\mu(x)\bar{\eta}_\nu(0)]\Big |0\Big\rangle.
	\end{equation}

In the zero-width resonance approximation, one can write the imaginary part of the correlation function in the rest frame $\vec{p}=0$ as (for the $g_{\mu\nu}$ and the $g_{\mu\nu}\pslash$ structures)
	\begin{align}\label{ppRHS}
		\begin{split}
		\text{Im}~\Pi^p(p_0)
		&=\sum_n \Big[(\lambda_n^+)^2 \frac{\gamma_0+1}{2}\delta(p_0-m_n^+)+(\lambda_n^-)^2 \frac{\gamma_0-1}{2}\delta(p_0-m_n^-)\Big]\\ 
		&\equiv \gamma_0 A(p_0)+B(p_0),
		\end{split}
	\end{align}
where $A(p_0)$ and $B(p_0)$ are defined as
	\begin{align}\label{funcAB}
		\begin{split}
		&A(p_0)=\frac{1}{2}\sum_n [(\lambda_n^+)^2 \delta(p_0-m_n^+)+(\lambda_n^-)^2 \delta(p_0-m_n^-)],\\
		&B(p_0)=\frac{1}{2}\sum_n [(\lambda_n^+)^2 \delta(p_0-m_n^+)-(\lambda_n^-)^2 \delta(p_0-m_n^-)],
		\end{split}
	\end{align}
and $\lambda_\pm=(2\pi)^2\tilde{\lambda}_\pm$. The parity-projected Delta sum rules are then given as
	\begin{align}\label{PPsr}
		\begin{split}
		&A(M,w_\pm)\pm B(M,w_\pm)=\\
		&\quad 2m_\pm\,(\lambda_{\pm})^2 \int^{w_\pm}_0 I(m_\pm,\Gamma_\pm)\,e^{-p_0^2/M^2}\,dp_0\, + 2m_\pm^\prime\,(\lambda_\pm^{\prime})^2 \int^{w_\pm}_0  I(m_\pm^\prime,\Gamma_\pm^\prime)\,e^{-p_0^2/M^2}\,dp_0,
		\end{split}
	\end{align}
with 
	\begin{align}\label{LHSAB}
		\begin{split}	
		&A(M,w_\pm)=\frac{1}{10}\,P^\pm_5\,L^{4/27}-\frac{5}{72}\,b \,P^\pm_1\,L^{24/27}+\frac{2}{3}\,\kappa\,a^2\,L^{28/27},\\	
		&B(M,w_\pm)=\frac{4}{3}\,a\,P^\pm_2\,L^{16/27} -\frac{2}{3}\,m_0^2\,a\,P^\pm_0\,L^{2/27},
		\end{split}
	\end{align}
where we define
	\begin{equation}\label{intfunc}
		P^\pm_n(M,w_\pm)\equiv\int^{w_\pm}_0\,dp_0\,p_0^n\,e^{-p_0^2/M^2}.
	\end{equation}
Note that the functions $A$ and $B$ can be readily identified from Eq.~(\ref{momsp}). In Eq.~(\ref{PPsr}), $m_+$ ($m_-$) and $\lambda_+$ ($\lambda_-$) denote the mass and the overlap amplitude of the positive-(negative-)parity Delta resonance, respectively, $w_+$ ($w_-$) is the continuum threshold for the positive-(negative-)parity sum rule and the primed notation refers to those parameters for the first positive-(negative-)parity excitation. We have also made use of the relation
	\begin{equation}\label{deltcon}
		\delta(p_0-m_\pm)\equiv 2m_\pm\,\delta(p_0^2-m_\pm^2)
	\end{equation}
on the positive $p_0$-axis and replaced the $\delta$-function with the Breit-Wigner form in Eq.~(\ref{BWf}) (taking $s\rightarrow p_0^2$). We have included the first Delta excitation with the second term on the right-hand side of the sum rule in Eq.~(\ref{PPsr}). The functions in Eq.~(\ref{LHSAB}) can be written more explicitly as~\cite{Lee:2006bu}
	\begin{align}
		\begin{split}	
		A(M,w_\pm)&=\frac{1}{10}\,\left[ \left(1-e^{-w_\pm^2/M^2} \right) M^6 - w_\pm^2 e^{-w_\pm^2/M^2}M^4-\frac{w_\pm^4}{2} e^{-w_\pm^2/M^2} M^2\right]\,L^{4/27}\\
		&-\frac{5}{144}\,b\,\left(1-e^{-w_\pm^2/M^2} \right)M^2 \,L^{24/27} +\frac{2}{3}\,\kappa\,a^2\,L^{28/27},\\	
		B(M,w_\pm)&=\frac{2}{3}\,a\,\left[I(w_\pm)-w_\pm e^{-w_\pm^2/M^2}\right] M^2\,L^{16/27} -\frac{2}{3}\,m_0^2\,a\,I(w_\pm)\,L^{2/27},
		\end{split}
	\end{align}
with \(I(w)=\int_0^w e^{-x^2/M^2}\,dx\).

%%%%%%%%%%%%%%%%%%% Subsection 3-b %%%%%%%%%%%%%%%%%%%%%%%%%%%%
\subsection{Numerical Analysis of the Parity-Projected Sum Rules}

We follow a similar procedure as above and first try a three-parameter fit to the LHS of Eq.~(\ref{PPsr}) for the positive-parity state including $m_+$, $\lambda_+$ and $w_+$, with the zero-width approximation and assuming that the first Delta excitation lies in the continuum. The valid Borel regions are estimated similarly as in the traditional sum rules: we take $0.95 \le M \le 1.30$~GeV. A three-parameter fit returns a continuum threshold value smaller than the resonance mass, which is an unphysical solution. This case, which signals that the OPE does not have enough information to resolve all the three parameters \emph{simultaneously}, has been realized before in various other works on QCDSR with Monte-Carlo fits~\cite{Lee:1996dc,Lee:1997ix,Lee:1997jk,Lee:2006bu}. In order to proceed then, we switch to a two-parameter fit by fixing the continuum threshold at a value suggested by the the Particle Data Group (PDG)~\cite{Yao:2006px} and the traditional sum rules above, as was also done in the previous works. To put it more explicitly, before fixing it we have an idea about the value of the continuum threshold from the traditional sum-rules analysis and experimental Delta spectrum. We fix the continuum threshold at a reasonable value accordingly.

At this point we would like to make a technical remark: the physical spectrum of the Delta resonance with a $p$-wave starts at $p_0=(m_N+m_\pi)$ while the tail of the infinite Breit-Wigner shape we utilize extends unphysically to low-$p_0$ region. In order to eliminate the overlap of the spectrum with the region that falls below the $\pi$-$N$ threshold, one can use a parametrization of the spectral function in terms of a $p_0$-dependent width (see Ref.~\cite{Leupold:1997dg} for the corresponding case of the $\rho$-meson). Such a constraint aggravates the numerical difficulties that arise from a multi-parameter fit. Therefore, we choose to use the simplest parametrization of the Breit-Wigner form with a constant width~\cite{Yao:2006px}, which serves best for our purposes. For the parity-projected sum rules, the part of the spectrum below the $\pi$-$N$ threshold amounts to 15\% of the total spectrum (for physical mass and the width). In the case of the traditional sum rules, where the physical spectrum starts at $s=(m_N+m_\pi)^2$, we have smaller overlap thanks to the four-pulse approximation in Eq.(\ref{unitpulse}): actually for OSR, which performs better as compared to ESR, all the rectangular pulses are above the $\pi$-$N$ threshold, while for ESR, only 9\% of the Breit-Wigner distribution lies below the threshold. 

\begin{figure}
	[t] 
	\includegraphics[scale=0.50]{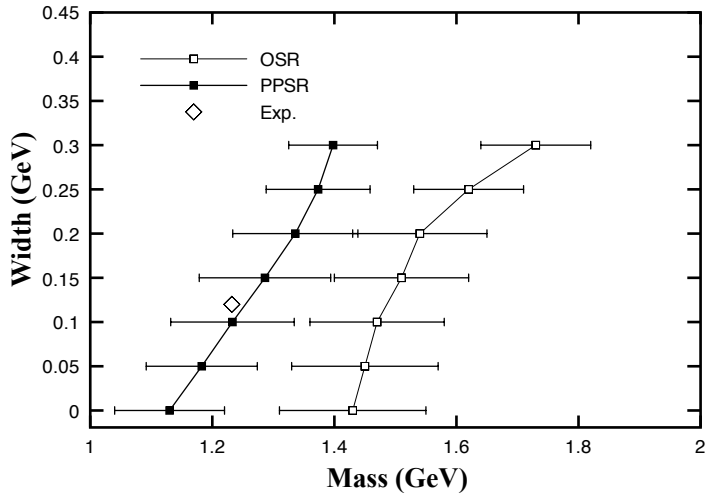} \caption{Correlation of the mass and the width of the $\Delta(1232)$ resonance using OSR (filled squares) and the parity-projected sum rules (PPSR) (empty squares). The diamond marks the experimental mass and the width of $\Delta(1232)$. Each datum point has been obtained from a consideration of 100 configurations. The error bars associated with each point stem from the uncertainties in the condensate values.} \label{masswidth}
\end{figure}

In Table~\ref{parproj}, the obtained values of the fit parameters from a consideration of 1000 parameter sets are presented, for two different values of the continuum threshold; $w_+=1.5$~GeV and $w_+=1.6$~GeV. In the zero-width approximation for $w_+=1.5$~GeV, we obtain $m_+=1.13 \pm 0.09$~GeV, whereas a two-parameter fit including $m_+$ and $\lambda_+$ with the finite-width effects gives $m_+=1.23 \pm 0.09$~GeV. For the latter, we evaluate the integrals over the Breit-Wigner shape numerically and obtain a Delta mass value in agreement with the experimental one. If we instead fix the continuum threshold at $w_+=1.6$~GeV, the mass values produced by the fit are changed by 10\%. Comparison of the mass values obtained with the zero-width approximation and with the finite-width effects suggests that the finite-width contributions are rather important for the parity-projected Delta sum rules. These effects lead to a change of $\sim$~20\% and $\sim$~10\% in $\lambda_+^2$ and $m_+$, respectively. A search for the width and the overlap amplitude in a fit with $m_+=1.232$~GeV successfully reproduces the experimental result; that is $\Gamma_+=0.118 \pm 0.002$~GeV~\cite{Yao:2006px}. The overlap amplitudes are compatible with those in Table~\ref{hadpar} from the traditional sum rules. In order to analyze the impact of the mass variation on the width, we allow a 10\% variation of the input ground and excited state masses. We observe that the uncertainties in the widths are increased by a mass variation. To further illustrate the correlation of the mass and the width of the $\Delta(1232)$ resonance, in Fig.(\ref{masswidth}) we present the results obtained with a broad range of mass and width using OSR and the parity-projected sum rule. As was shown in Section~\ref{sect2}, OSR performs better as compared to ESR, however overestimates the mass value. It is seen from the figure that the masses start to deviate from their initial values towards higher width region for both sum rules. It is clear that the mass increases with the increasing width and the lowest mass is given by the calculation of the spectral function with a $\delta$-function representation. This observation is in agreement with that of Ref.~\cite{Leupold:1997dg} in the case of the $\rho$-meson.

\begin{table}[t]
	\addtolength{\tabcolsep}{3pt}
	 \caption{The obtained values of the parameters representing the positive-parity $\Delta(1232)$ and $\Delta(1600)$ resonances from a consideration of 1000 parameter sets using the parity-projected sum rule in Eq.~(\ref{PPsr}). For the excited $\Delta(1600)$ resonance, the mass and the width of the lowest-lying $\Delta(1232)$ resonance are fixed at $m_+=1.232$~GeV and $\Gamma_+=0.12$~GeV, respectively. The Borel window is taken as $0.95 \le M \le 1.30$~GeV when we consider only the $\Delta(1232)$ resonance, while the window is broadened to $0.95 \le M \le 1.50$~GeV when the $\Delta(1600)$ resonance is included. The fixed parameters are denoted by an asterisk in each case and the numbers inside brackets are experimental values as given by PDG~\cite{Yao:2006px}.} \centering 
	\begin{tabular}{ccccc}
		\hline\hline $\Delta(1232)$ & $\Gamma_+$~(GeV) & $m_+$~(GeV) & $w_+$~(GeV) & $\lambda_+^2$(GeV$^6$) \\
		\hline  & $[0.118 \pm 0.002]$ & $[1.232]$ & & \\
		 & $0~^\ast$ & $1.13 \pm 0.09$ & $1.5~^\ast$ & $1.26 \pm 0.19$ \\[-1ex]
		 &  $0~^\ast$ & $1.19 \pm 0.09$ & $1.6~^\ast$ & $1.57 \pm 0.20$ \\[1ex]
		&   $0.12~^\ast$ & $1.23 \pm 0.09$ & $1.5~^\ast$ &$1.48 \pm 0.19$ \\[-1ex]
		&   $0.12~^\ast$ & $1.31 \pm 0.09$ & $1.6~^\ast$ &$1.83 \pm 0.23$ \\[1ex]
		 &  $0.14\pm 0.08$  & $1.232~^\ast$ & $1.5~^\ast$ &$1.46 \pm 0.28$ \\[-1ex]
		 &  $0.10\pm 0.06$  & $1.232~^\ast$ & $1.6~^\ast$ &$1.98 \pm 0.47$ \\[-1ex]
		&  $0.15\pm 0.10$  & $1.23 \pm 0.12~^\ast$ & $1.5~^\ast$ &$1.55 \pm 0.49$ \\[1ex]
		\hline  $\Delta(1600)$ & $\Gamma^\prime_+$~(GeV) & $m^\prime_+$~(GeV) & $w^\prime_+$~(GeV) & $(\lambda^\prime_+)^2$(GeV$^6$) \\
		\hline  & $[0.35 \pm 0.10]$ & $[1.625 \pm 0.075]$ & & \\
		& $0.22 \pm 0.07$ & $1.6~^\ast$ & $1.8~^\ast$ & $3.47 \pm 0.45$ \\[-1ex]
		&  $0.20 \pm 0.09$ & $1.6 \pm 0.1~^\ast$ & $1.8~^\ast$ & $3.14 \pm 0.87$\\[-1ex]
		&  $0.20 \pm 0.08$ & $1.6~^\ast$ & $1.85 \pm 0.05~^\ast$ & $3.42 \pm 0.73$ \\[1ex]
		\hline\hline
	\end{tabular}
	\label{parproj}
\end{table}

\begin{table}[t]
	 \caption{Same as Table~\ref{parproj} but for the negative-parity $\Delta(1700)$ and $\Delta(1940)$ resonances. For the excited $\Delta(1940)$ resonance, the mass and the width of the lowest-lying $\Delta(1700)$ resonance are fixed at $m_-=1.7$~GeV and $\Gamma_-=0.3$~GeV, respectively. The Borel window is taken as $1.30 \le M \le 1.55$~GeV.} \centering 
	\begin{tabular}{ccccc}
		\hline\hline $\Delta(1700)$ & $\Gamma_-$~(GeV) & $m_-$~(GeV) & $w_-$~(GeV$^6$) & $\lambda_-^2$ \\
		\hline  & $[0.30 \pm 0.10]$ & $[1.710 \pm 0.040]$ & & \\
		& $0~^\ast$ & $1.45 
		\pm 0.67$ & $2.4~^\ast$ & $0.68 \pm 0.41$ \\[-1ex]
		 &   $0.30~^\ast$  & $1.57 \pm 0.32$ &  &$0.95 \pm 0.29$\\[-1ex]
		 &   $0.42\pm 0.36$  & $1.70~^\ast$ &  &$0.95 \pm 0.18$\\
		\hline $\Delta(1940)$ & $\Gamma_-^\prime$~(GeV) & $m_-^\prime$~(GeV) & $w_-^\prime$~(GeV$^6$) & $(\lambda_-^\prime)^2$ \\
		\hline  & $[?]$ & $[\approx 1.940]$ & & \\
		 & $0.65 \pm 0.29$ & 
		$1.94~^\ast$ & $2.4~^\ast$ & $1.41 \pm 0.88$\\
		\hline\hline
	\end{tabular}
	\label{parprojneg}
\end{table}

In order to include the first positive-parity excitation in our analysis, we apply a three-parameter fit by taking $m_+=1.232$~GeV, $\Gamma_+=0.12$~GeV and $m_+^\prime=1.6$~GeV. In this case, the upper limit on the valid Borel window is increased according to the selection criterion above, since the continuum contributes less with the increasing threshold: we take $0.95 \le M \le 1.50$~GeV. In Table~\ref{parproj}, we present the width values of the first positive-parity excitation obtained from the fit, for different values of the resonance mass and the continuum threshold. We also allow a reasonable variation in the input mass and continuum threshold parameters, by generating them via Monte Carlo. The resulting width values are in agreement with the experimental result, which is $\Gamma_+^\prime=0.35 \pm 0.10$~GeV as given by PDG~\cite{Yao:2006px}. A comparison of the overlap amplitude values produced by the fit indicates that the lowest-lying and the first excited states couple to the interpolating current with strengths of the same order.

In this work, we have also analyzed the negative-parity Delta states; however, the scarcity of the experimental information for the negative-parity resonances prevents us from constraining the parameters, which represent these resonances. Particle Data Group (PDG) lists the lowest-lying negative-parity ($J=3/2$) state as $\Delta(1700)$ with Breit-Wigner width of $0.3 \pm 0.1$~GeV. Assuming that the first Delta excitation lies in the continuum, we have first applied a two-parameter fit to the LHS of Eq.~(\ref{PPsr}) including $m_-$ and $\lambda_-$ and have found that the experimental mass value of the $\Delta(1700)$ resonance can be obtained only for $w_-\geq 2.4$~GeV (see Table~\ref{parprojneg}). Fixing the continuum threshold value at $w_-=2.4$~GeV, we obtain $m_-=1.45\pm 0.67$~GeV with the zero-width approximation. If we adopt the average value for the width of the $\Delta(1700)$ resonance as $\Gamma_-=0.3$~GeV~\cite{Yao:2006px} and apply a fit with finite-width effects, we obtain $m_-=1.57\pm 0.32$~GeV, while a search for the width with $m_-=1.7$~GeV produces $\Gamma_-=0.42 \pm 0.36$~GeV. The relatively high value, at which we have to fix the continuum threshold in order to obtain the experimental mass value of the $\Delta(1700)$ resonance suggests that the excited-state contributions may not be embedded in the continuum and therefore the QCDSR analysis cannot be limited to the lowest-lying resonance. The first negative-parity excitation (with $J=3/2$) is listed as $\Delta(1940)$ by PDG and the available data for the Breit-Wigner width of this resonance have large uncertainties ($0.10\sim 0.78$~GeV). To continue, we include in our fit the $\Delta(1940)$ resonance with $m_-^\prime=1.94$~GeV. We apply a three-parameter fit by fixing the continuum threshold value at $w_-=2.4$~GeV and by taking $m_-=1.7$~GeV, $\Gamma_-=0.3$~GeV. As a result of this fit we obtain $\Gamma_-^\prime=0.65 \pm 0.29$~GeV, which favors a large value for the width of $\Delta(1940)$. Despite the naive analysis as a result of the large uncertainties in the phenomenological input, our results for the negative-parity Delta resonance parameters are consistent with those given by PDG~\cite{Yao:2006px}. 

%%%%%%%%%%%%%%%%%%% Section 3 %%%%%%%%%%%%%%%%%%%%%%%%%%%%
\section{Conclusion and Discussions} \label{sect4}
In summary, we have analyzed the finite-width effects on the QCD sum rules of the Delta baryons, through explicit utilization of the Breit-Wigner shape. We have observed that although the finite-width effects give minor contributions to the traditional sum rules (about $3\%$ change in the $\Delta(1232)$ mass value), the effects on the parity-projected sum rules are rather significant (about $10\%$ change in the $\Delta(1232)$ mass value). The deviations from the narrow-resonance approximation in QCDSR analysis leads to results in better agreement with the experiment for the parity-projected case. Our findings for the width of the $\Delta(1232)$ resonance from a Monte Carlo analysis of the parity-projected sum rule are in agreement with the experimental results. The first positive-parity Delta excitation, $\Delta(1600)$, has also been considered as a sub-continuum resonance using parity-projected QCDSR, and its Breit-Wigner width has been calculated, where the experimental width value has been produced. For both resonances, in order to analyze the mass-width correlation, we allowed a variation in the input mass and the width values. We have found that while the masses are insensitive on a width variation for small width values, the uncertainties in the widths are increased with respect to a mass variation. This in turn means that the boundaries are more restrictive for the mass and less for the width. Our numerical analysis shows that the masses increase with the increasing widths and the lowest mass is given by the calculation of the spectral function with a $\delta$-function representation. For the negative-parity case, large experimental uncertainties prevent us from constraining the input parameters, therefore only a naive analysis can be made. A reasonable agreement with the experimental values has been obtained though, which favors a large value for the Breit-Wigner width of the $\Delta(1940)$ resonance. It is desirable to extend this analysis to all decuplet baryons and their excited states. A work along this line is still in progress. 

\acknowledgments
Discussions with G. Turan and D. Jido are gratefully acknowledged. This work has been supported by the Japan Society for the Promotion of Science under contract number P06327.


\begin{thebibliography}{24}
\expandafter\ifx\csname natexlab\endcsname\relax\def\natexlab#1{#1}\fi
\expandafter\ifx\csname bibnamefont\endcsname\relax
  \def\bibnamefont#1{#1}\fi
\expandafter\ifx\csname bibfnamefont\endcsname\relax
  \def\bibfnamefont#1{#1}\fi
\expandafter\ifx\csname citenamefont\endcsname\relax
  \def\citenamefont#1{#1}\fi
\expandafter\ifx\csname url\endcsname\relax
  \def\url#1{\texttt{#1}}\fi
\expandafter\ifx\csname urlprefix\endcsname\relax\def\urlprefix{URL }\fi
\providecommand{\bibinfo}[2]{#2}
\providecommand{\eprint}[2][]{\url{#2}}

\bibitem[{\citenamefont{Ioffe}(1981)}]{Ioffe:1981kw}
\bibinfo{author}{\bibfnamefont{B.~L.} \bibnamefont{Ioffe}},
  \bibinfo{journal}{Nucl. Phys.} \textbf{\bibinfo{volume}{B188}},
  \bibinfo{pages}{317} (\bibinfo{year}{1981}).

\bibitem[{\citenamefont{Belyaev and Ioffe}(1982)}]{Belyaev:1982sa}
\bibinfo{author}{\bibfnamefont{V.~M.} \bibnamefont{Belyaev}} \bibnamefont{and}
  \bibinfo{author}{\bibfnamefont{B.~L.} \bibnamefont{Ioffe}},
  \bibinfo{journal}{Sov. Phys. JETP} \textbf{\bibinfo{volume}{56}},
  \bibinfo{pages}{493} (\bibinfo{year}{1982}).

\bibitem[{\citenamefont{Reinders et~al.}(1985)\citenamefont{Reinders,
  Rubinstein, and Yazaki}}]{Reinders:1984sr}
\bibinfo{author}{\bibfnamefont{L.~J.} \bibnamefont{Reinders}},
  \bibinfo{author}{\bibfnamefont{H.}~\bibnamefont{Rubinstein}},
  \bibnamefont{and} \bibinfo{author}{\bibfnamefont{S.}~\bibnamefont{Yazaki}},
  \bibinfo{journal}{Phys. Rept.} \textbf{\bibinfo{volume}{127}},
  \bibinfo{pages}{1} (\bibinfo{year}{1985}).

\bibitem[{\citenamefont{Kodama and Oka}(1994)}]{Kodama:1994pn}
\bibinfo{author}{\bibfnamefont{N.}~\bibnamefont{Kodama}} \bibnamefont{and}
  \bibinfo{author}{\bibfnamefont{M.}~\bibnamefont{Oka}},
  \bibinfo{journal}{Phys. Lett.} \textbf{\bibinfo{volume}{B340}},
  \bibinfo{pages}{221} (\bibinfo{year}{1994}).

\bibitem[{\citenamefont{Hwang and Yang}(1994)}]{Hwang:1994vp}
\bibinfo{author}{\bibfnamefont{W.~Y.~P.} \bibnamefont{Hwang}} \bibnamefont{and}
  \bibinfo{author}{\bibfnamefont{K.-C.} \bibnamefont{Yang}},
  \bibinfo{journal}{Phys. Rev.} \textbf{\bibinfo{volume}{D49}},
  \bibinfo{pages}{460} (\bibinfo{year}{1994}).

\bibitem[{\citenamefont{Lee}(1998{\natexlab{a}})}]{Lee:1997ix}
\bibinfo{author}{\bibfnamefont{F.~X.} \bibnamefont{Lee}},
  \bibinfo{journal}{Phys. Rev.} \textbf{\bibinfo{volume}{C57}},
  \bibinfo{pages}{322} (\bibinfo{year}{1998}{\natexlab{a}}).

\bibitem[{\citenamefont{Lee}(2007)}]{Lee:2006bu}
\bibinfo{author}{\bibfnamefont{F.~X.} \bibnamefont{Lee}},
  \bibinfo{journal}{Nucl. Phys.} \textbf{\bibinfo{volume}{A791}},
  \bibinfo{pages}{352} (\bibinfo{year}{2007}).

\bibitem[{\citenamefont{Erkol and Oka}(2008)}]{Erkol:2007sj}
\bibinfo{author}{\bibfnamefont{G.}~\bibnamefont{Erkol}} \bibnamefont{and}
  \bibinfo{author}{\bibfnamefont{M.}~\bibnamefont{Oka}},
  \bibinfo{journal}{Phys. Lett.} \textbf{\bibinfo{volume}{B659}},
  \bibinfo{pages}{176} (\bibinfo{year}{2008}).

\bibitem[{\citenamefont{Shifman
  et~al.}(1979{\natexlab{a}})\citenamefont{Shifman, Vainshtein, and
  Zakharov}}]{Shifman:1978bx}
\bibinfo{author}{\bibfnamefont{M.~A.} \bibnamefont{Shifman}},
  \bibinfo{author}{\bibfnamefont{A.~I.} \bibnamefont{Vainshtein}},
  \bibnamefont{and} \bibinfo{author}{\bibfnamefont{V.~I.}
  \bibnamefont{Zakharov}}, \bibinfo{journal}{Nucl. Phys.}
  \textbf{\bibinfo{volume}{B147}}, \bibinfo{pages}{385}
  (\bibinfo{year}{1979}{\natexlab{a}}).

\bibitem[{\citenamefont{Shifman
  et~al.}(1979{\natexlab{b}})\citenamefont{Shifman, Vainshtein, and
  Zakharov}}]{Shifman:1978by}
\bibinfo{author}{\bibfnamefont{M.~A.} \bibnamefont{Shifman}},
  \bibinfo{author}{\bibfnamefont{A.~I.} \bibnamefont{Vainshtein}},
  \bibnamefont{and} \bibinfo{author}{\bibfnamefont{V.~I.}
  \bibnamefont{Zakharov}}, \bibinfo{journal}{Nucl. Phys.}
  \textbf{\bibinfo{volume}{B147}}, \bibinfo{pages}{448}
  (\bibinfo{year}{1979}{\natexlab{b}}).

\bibitem[{\citenamefont{Leupold et~al.}(1998)\citenamefont{Leupold, Peters, and
  Mosel}}]{Leupold:1997dg}
\bibinfo{author}{\bibfnamefont{S.}~\bibnamefont{Leupold}},
  \bibinfo{author}{\bibfnamefont{W.}~\bibnamefont{Peters}}, \bibnamefont{and}
  \bibinfo{author}{\bibfnamefont{U.}~\bibnamefont{Mosel}},
  \bibinfo{journal}{Nucl. Phys.} \textbf{\bibinfo{volume}{A628}},
  \bibinfo{pages}{311} (\bibinfo{year}{1998}).

\bibitem[{\citenamefont{Iqbal et~al.}(1996{\natexlab{a}})\citenamefont{Iqbal,
  Jin, and Leinweber}}]{Iqbal:1995gp}
\bibinfo{author}{\bibfnamefont{M.~J.} \bibnamefont{Iqbal}},
  \bibinfo{author}{\bibfnamefont{X.-m.} \bibnamefont{Jin}}, \bibnamefont{and}
  \bibinfo{author}{\bibfnamefont{D.~B.} \bibnamefont{Leinweber}},
  \bibinfo{journal}{Phys. Lett.} \textbf{\bibinfo{volume}{B367}},
  \bibinfo{pages}{45} (\bibinfo{year}{1996}{\natexlab{a}}).

\bibitem[{\citenamefont{Iqbal et~al.}(1996{\natexlab{b}})\citenamefont{Iqbal,
  Jin, and Leinweber}}]{Iqbal:1995qc}
\bibinfo{author}{\bibfnamefont{M.~J.} \bibnamefont{Iqbal}},
  \bibinfo{author}{\bibfnamefont{X.-m.} \bibnamefont{Jin}}, \bibnamefont{and}
  \bibinfo{author}{\bibfnamefont{D.~B.} \bibnamefont{Leinweber}},
  \bibinfo{journal}{Phys. Lett.} \textbf{\bibinfo{volume}{B386}},
  \bibinfo{pages}{55} (\bibinfo{year}{1996}{\natexlab{b}}).

\bibitem[{\citenamefont{Elias et~al.}(1997)\citenamefont{Elias, Fariborz,
  Samuel, Shi, and Steele}}]{Elias:1997ya}
\bibinfo{author}{\bibfnamefont{V.}~\bibnamefont{Elias}},
  \bibinfo{author}{\bibfnamefont{A.}~\bibnamefont{Fariborz}},
  \bibinfo{author}{\bibfnamefont{M.~A.} \bibnamefont{Samuel}},
  \bibinfo{author}{\bibfnamefont{F.}~\bibnamefont{Shi}}, \bibnamefont{and}
  \bibinfo{author}{\bibfnamefont{T.~G.} \bibnamefont{Steele}},
  \bibinfo{journal}{Phys. Lett.} \textbf{\bibinfo{volume}{B412}},
  \bibinfo{pages}{131} (\bibinfo{year}{1997}).

\bibitem[{\citenamefont{Jido and Oka}(1996)}]{Jido:1996zw}
\bibinfo{author}{\bibfnamefont{D.}~\bibnamefont{Jido}} \bibnamefont{and}
  \bibinfo{author}{\bibfnamefont{M.}~\bibnamefont{Oka}} (\bibinfo{year}{1996}),
  \eprint{hep-ph/9611322}.

\bibitem[{\citenamefont{Jido et~al.}(1996)\citenamefont{Jido, Kodama, and
  Oka}}]{Jido:1996ia}
\bibinfo{author}{\bibfnamefont{D.}~\bibnamefont{Jido}},
  \bibinfo{author}{\bibfnamefont{N.}~\bibnamefont{Kodama}}, \bibnamefont{and}
  \bibinfo{author}{\bibfnamefont{M.}~\bibnamefont{Oka}},
  \bibinfo{journal}{Phys. Rev.} \textbf{\bibinfo{volume}{D54}},
  \bibinfo{pages}{4532} (\bibinfo{year}{1996}).

\bibitem[{\citenamefont{Leinweber}(1997)}]{Leinweber:1995fn}
\bibinfo{author}{\bibfnamefont{D.~B.} \bibnamefont{Leinweber}},
  \bibinfo{journal}{Annals Phys.} \textbf{\bibinfo{volume}{254}},
  \bibinfo{pages}{328} (\bibinfo{year}{1997}).

\bibitem[{\citenamefont{Dominguez}(1984)}]{Dominguez:1984eh}
\bibinfo{author}{\bibfnamefont{C.~A.} \bibnamefont{Dominguez}},
  \bibinfo{journal}{Z. Phys.} \textbf{\bibinfo{volume}{C26}},
  \bibinfo{pages}{269} (\bibinfo{year}{1984}).

\bibitem[{\citenamefont{Chung et~al.}(1984)\citenamefont{Chung, Dosch, Kremer,
  and Schall}}]{Chung:1984gr}
\bibinfo{author}{\bibfnamefont{Y.}~\bibnamefont{Chung}},
  \bibinfo{author}{\bibfnamefont{H.~G.} \bibnamefont{Dosch}},
  \bibinfo{author}{\bibfnamefont{M.}~\bibnamefont{Kremer}}, \bibnamefont{and}
  \bibinfo{author}{\bibfnamefont{D.}~\bibnamefont{Schall}},
  \bibinfo{journal}{Z. Phys.} \textbf{\bibinfo{volume}{C25}},
  \bibinfo{pages}{151} (\bibinfo{year}{1984}).

\bibitem[{\citenamefont{Yao et~al.}(2006)}]{Yao:2006px}
\bibinfo{author}{\bibfnamefont{W.~M.} \bibnamefont{Yao}} \bibnamefont{et~al.},
  \bibinfo{journal}{J. Phys.} \textbf{\bibinfo{volume}{G33}},
  \bibinfo{pages}{1} (\bibinfo{year}{2006}).

\bibitem[{\citenamefont{Chu et~al.}(1993)\citenamefont{Chu, Grandy, Huang, and
  Negele}}]{Chu:1993cn}
\bibinfo{author}{\bibfnamefont{M.~C.} \bibnamefont{Chu}},
  \bibinfo{author}{\bibfnamefont{J.~M.} \bibnamefont{Grandy}},
  \bibinfo{author}{\bibfnamefont{S.}~\bibnamefont{Huang}}, \bibnamefont{and}
  \bibinfo{author}{\bibfnamefont{J.~W.} \bibnamefont{Negele}},
  \bibinfo{journal}{Phys. Rev.} \textbf{\bibinfo{volume}{D48}},
  \bibinfo{pages}{3340} (\bibinfo{year}{1993}).

\bibitem[{\citenamefont{Schafer et~al.}(1994)\citenamefont{Schafer, Shuryak,
  and Verbaarschot}}]{Schafer:1993ra}
\bibinfo{author}{\bibfnamefont{T.}~\bibnamefont{Schafer}},
  \bibinfo{author}{\bibfnamefont{E.~V.} \bibnamefont{Shuryak}},
  \bibnamefont{and} \bibinfo{author}{\bibfnamefont{J.~J.~M.}
  \bibnamefont{Verbaarschot}}, \bibinfo{journal}{Nucl. Phys.}
  \textbf{\bibinfo{volume}{B412}}, \bibinfo{pages}{143} (\bibinfo{year}{1994}).

\bibitem[{\citenamefont{Lee et~al.}(1997)\citenamefont{Lee, Leinweber, and
  Jin}}]{Lee:1996dc}
\bibinfo{author}{\bibfnamefont{F.~X.} \bibnamefont{Lee}},
  \bibinfo{author}{\bibfnamefont{D.~B.} \bibnamefont{Leinweber}},
  \bibnamefont{and} \bibinfo{author}{\bibfnamefont{X.-m.} \bibnamefont{Jin}},
  \bibinfo{journal}{Phys. Rev.} \textbf{\bibinfo{volume}{D55}},
  \bibinfo{pages}{4066} (\bibinfo{year}{1997}).

\bibitem[{\citenamefont{Lee}(1998{\natexlab{b}})}]{Lee:1997jk}
\bibinfo{author}{\bibfnamefont{F.~X.} \bibnamefont{Lee}},
  \bibinfo{journal}{Phys. Rev.} \textbf{\bibinfo{volume}{D57}},
  \bibinfo{pages}{1801} (\bibinfo{year}{1998}{\natexlab{b}}).

\end{thebibliography}
\end{document}